\documentclass[10pt,twocolumn]{article}
\usepackage{graphicx}
\usepackage{amsmath}
\usepackage{ol}

\begin{document}
\twocolumn[
\title{Reflectionless evanescent-wave amplification by two
dielectric planar waveguides}
\author{Mankei Tsang and Demetri Psaltis}
\date{\today}
\affiliation{
Department of Electrical Engineering, 
California Institute of Technology, Pasadena, California 91125}
\begin{abstract}
Utilizing the underlying physics of evanescent-wave amplification by
a negative-refractive-index slab, it is shown that evanescent
waves with specific spatial frequencies can also be amplified without any
reflection simply by two dielectric planar waveguides.  The simple
configuration allows one to take advantage of the high resolution
limit of a high-refractive-index material without contact with the
object.
\end{abstract}
]
\maketitle

Conventional optical imaging systems cannot resolve features smaller
than the optical wavelength, because the high-spatial-frequency modes
that describe the subwavelength features are evanescent waves, which
exponentially decay away from the object and do not propagate to the
far field. Observing the evanescent waves is therefore one of the most
important yet formidable challenges in the field of optics, with
important applications in optical lithography, data storage, and
microscopy. Near-field scanning optical microscopy can detect the
evanescent waves \cite{betzig}, but it requires scanning, which may
not be desirable for many applications.  A groundbreaking proposal by
Pendry suggests that evanescent waves can be amplified without
any reflection in a negative-refractive-index slab \cite{pendry},
causing significant interest as well as controversy \cite{controversy}
in the mechanism of evanescent-wave amplification (EWA). On the
practical side, the fabrication of a negative-refractive-index
material for optical frequencies is a challenging task, as it requires
both negative permittivity and negative permeability, the latter of
which does not naturally occur in materials, and methods of
implementing an effective negative refractive index
\cite{zhang,zhang2,shalaev} often introduce significant loss
detrimental to the EWA process.  As proposed by Pendry \cite{pendry}
and experimentally demonstrated by Fang \textit{et al.}\ \cite{fang},
a negative permittivity in a metal slab can also amplify evanescent
waves to some extent, but the thickness of the slab is limited by the
electrostatic approximation as well as loss. A simpler EWA scheme that
utilizes less lossy materials would thus be desirable.

Along this direction, Luo \textit{et al.}\ propose that a photonic
crystal slab can be used to amplify evanescent waves \cite{luo}, since
evanescent waves with specific spatial frequencies can be coupled into
the bound states of the photonic crystal slab, and the build-up of the
bound states produces an enhanced evanescent tail on the other side of
the slab. Apart from the difficulty in fabricating a three-dimensional
photonic crystal for two-dimensional imaging, the kind of EWA achieved
by a photonic crystal slab is not ideal, because the build-up of the
bound states also creates enhanced reflected evanescent waves, causing
multiple evanescent wave reflections between the object and the
photonic crystal. On the other hand, in order to obtain information
about the output evanescent waves on the image plane, energy must be
extracted, and the only way for the detector to ``tell'' the imaging
system to give up energy is via a reflected evanescent wave. In other
words, detection of an evanescent wave always creates a reflected
evanescent wave, so there exist multiple reflections between an
imaging system and the detector as well.  Since the magnitudes of
evanescent wave transmission and reflection coefficients can be larger
than 1 or even infinite, multiple evanescent wave reflections can be
very significant and should not be ignored in the design of near-field
imaging systems.  An ideal near-field imaging system should hence have
unit transmission as well as zero reflection, \emph{as if the imaging
system is not there and the object directly touches the image
plane}. This ideal behavior also allows multiple imaging systems to be
cascaded and a longer distance between the object and the detector.

In this Letter, the underlying physics of reflectionless
evanescent-wave amplification (REWA) by the use of a
negative-refractive-index slab is explained, and, using this
knowledge, it is shown that evanescent waves with specific spatial
frequencies can be amplified without reflection simply by two
dielectric planar waveguides.
Since loss in a dielectric can be orders-of-magnitude lower than
metals or metamaterials, our proposed scheme is the simplest way of
experimentally demonstrating the intriguing phenomenon of REWA and
offers simple alternatives to the use of left-handed materials,
surface plasmons, or photonic crystals for near-field imaging
applications.

One of the most poorly understood aspects of Pendry's proposal is that
at the interface of an $n = 1$ material and an $n = -1$ material, the
transmission and reflection coefficients are theoretically infinite
\cite{pendry}. Mathematically this indicates the presence of an
unstable pole on the imaginary axis in the complex
transverse-spatial-frequency ($s = ik_x$) plane, and physically the
transmitted and reflected evanescent optical fields must therefore
increase linearly along a semi-infinite interface. This is hardly
surprising if one recalls the well-known fact that infinite scattering
coefficients correspond to bound-state solutions, so the incoming
evanescent waves are simply resonantly-coupled into the waveguide
modes of the interface. The most peculiar aspect of Pendry's interface
is that the scattering coefficients are always infinite, meaning that
bound-state solutions exist for all $k_x$. This is not true for other
waveguides, including photonic crystals \cite{luo}, which have
discrete bound states with different discrete $k_x$'s.  In particular,
for ideal surface plasmons, only one bound state exists.

First, consider a dielectric slab with thickness $a$ and
refractive index $n_1$ in the $x-y$ plane.  Suppose that an evanescent
s-polarized wave with an electric field exponentially decaying along
the positive $z$ axis, given by $\mathbf{E}_{0+} = [0, 1, 0]\exp(ik_z z+ik_x
x-i\omega t)$, impinges on the slab, where $k_x$ is assumed to have
subwavelength resolution, so $k_x > \omega n_0/c$, $k_z$ is determined
by the dispersion relation, given by $k_z= i\sqrt{k_x^2-k_0^2}$, $k_0
= \omega n_0/c$, and $n_0$ is the refractive index of the
surroundings.  Considering the first interface between $n_0$ and $n_1$
only, the reflected wave is $r[0, 1, 0]\exp(-ik_z
z+ik_x x -i\omega t)$, and the transmitted wave inside the slab is
$t[0, 1, 0]\exp(ik_z' z+ik_x x-i\omega t)$.  $k_x$
is the same on both sides of the interface, and $k_z'$ is given by the
dispersion relation $k_z' = \sqrt{k_1^2-k_x^2}$, where $k_1 = \omega
n_1/c$.  $k_z'$ is hereafter assumed to be real for waveguide modes to
exist. This restricts $k_x$ to be bounded by the wave numbers in the
two media,
\begin{align}
k_0< k_x < k_1.
\label{resolutionlimit}
\end{align}
The transmission and reflection coefficients across the first interface
are given by $t = 2k_z/(k_z+k_z')$ and
$r = (k_z-k_z')/(k_z+k_z')$ respectively.
Likewise, the scattering coefficients across the second interface are
$t' = 2k_z'/(k_z'+k_z)$ and
$r' = (k_z'-k_z)/(k_z' + k_z)$.
To obtain the total transmission, $\tau$, across the slab,
multiple scattering events must be summed,
\begin{align}
\tau &= t\exp(ik_z'a)t' + t\exp(ik_z' a)[r'\exp(ik_z'a)]^2t'+...\\
&=\frac{tt'\exp(ik_z'a)}{1-r'^2\exp(2ik_z'a)}.
\end{align}
The total reflection coefficient can be obtained similarly,
\begin{align}
\Gamma &= r + \frac{tt'r'\exp(2ik_z'a)}{1-r'^2\exp(2ik_z'a)}.
\end{align}
Waveguide modes correspond to those with evanescent tails
exponentially decaying away from the waveguide. In other words, the
total transmitted evanescent wave and the total reflected evanescent
wave for the waveguide modes can exist by themselves without any
incoming wave $\mathbf{E}_{0+}$, or, mathematically speaking, $\tau$
and $\Gamma$ are infinity. This happens when
\begin{align}
1-r'^2\exp(2ik_z'a) &= 0,
\label{pole}
\end{align}
which simply states that the accumulated phase in a round trip inside
the waveguide must be multiples of $2\pi$.  As both $k_z$ and $k_z'$
depend on $k_x$, Eq.~(\ref{pole}) is an eigenvalue equation of $k_x$
for the TE modes of the single waveguide.  A simple dielectric slab
can hence achieve EWA due to the waveguide mode coupling resonances,
similar to a photonic crystal \cite{luo}. If only subwavelength features
are concerned and all-angle negative refraction \cite{luo2} is not needed,
a complicated structure such as photonic crystal is not
necessary. However, just like a photonic crystal, the reflection
coefficient $\Gamma$ of a slab waveguide is also infinite, causing
potential problems with multiple reflections.

In Pendry's proposal, both interfaces of a negative-refractive-index slab
need to be considered for ideal REWA.  The two interfaces can be
considered as two waveguides, and the total transmission of the slab
exponentially increases with respect to the thickness of the slab, or
the distance between the two waveguides, when the single-interface
scattering coefficients are infinite. This suggests that REWA may also
exist for other kinds of double-waveguide structures, when the
resonant coupling condition of the single waveguide is reached.

\begin{figure}[htbp]
\centerline{\includegraphics[width=0.4\textwidth]{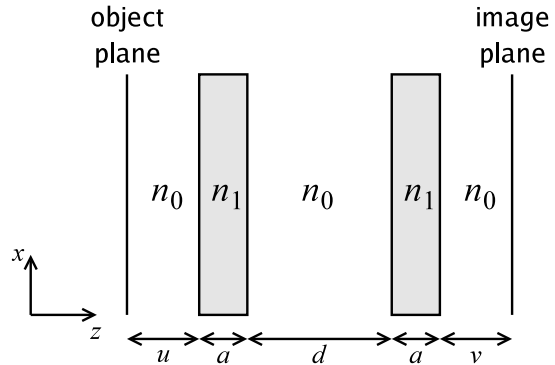}}
\caption{Reflectionless evanescent-wave amplification (REWA) by two
slab waveguides, where $n_1 > n_0$.}
\label{twoslabs}
\end{figure}

Now let us go back to the dielectric slab waveguide example and add
another identical waveguide a distance $d$ away from the first, as
depicted in Fig.~\ref{twoslabs}. The total transmission coefficient
for this double-waveguide structure is
\begin{align}
T &= \frac{\tau^2\exp(ik_z d)}{1-\Gamma^2\exp(2ik_z d)}.
\end{align}
When $k_x$ coincides with one of the single-waveguide bound-state
eigenvalues determined by Eq.~(\ref{pole}), the total transmission becomes
\begin{align}
\lim_{r'^2\exp(2ik_z'a)\to 1}T &= 
-\exp(-ik_z d),
\label{T2}
\end{align}
which increases exponentially with respect to $d$.
The total reflection coefficient of the double-waveguide structure is
likewise given by
$R =\Gamma + \tau^2\Gamma\exp(2ik_z d)/[1-\Gamma^2\exp(2ik_zd)]$,
and in the limit of $k_x$ being a bound-state eigenvalue of a single
waveguide,
\begin{align}
\lim_{r'^2\exp(2ik_z'a)\to 1}R
= 0.
\label{R2}
\end{align}
Hence, an evanescent wave can propagate with perfect transmission and
zero reflection in the setup depicted in Fig.~\ref{twoslabs}, thereby
achieving REWA, if $u+v=d$ and the resonant single-waveguide coupling
condition is reached. Identical results can also be derived for
p-polarized waves and TM modes. REWA should be quite general for any
kind of symmetric and identical waveguides,
so two photonic crystal slabs may be
used to achieve all-angle negative refraction \cite{luo2} and REWA
simultaneously.

For imaging applications, it is important to stress that the
double-waveguide device only beats the resolution limit of the
cladding layer with refractive index $n_0$, but not the resolution
limit of the core layer with refractive index $n_1$. This is because
the bound-state eigenvalues of $k_x$ are bounded by wave numbers of
the two media, as shown by Eq.~(\ref{resolutionlimit}).  That said, a
waveguide can be designed such that the maximum $k_x$ of a waveguide
mode is close to the wave number of the core medium, so the proposed
device can still take advantage of the high resolution limit offered
by a high-refractive-index material without contact with the
object. This can be advantageous for many applications because many
solids have higher refractive indices than fluids but it is not very
practical to fill the whole imaging system with solids as in oil
immersion microscopy. Furthermore, for biomedical imaging
applications, it is not always possible to place the
high-refractive-index material directly in touch with the object
plane, as the contact may damage the biological sample, or one may
desire to put the object plane inside a semi-transparent object, such
as a cell.

Promising high-refractive-index material candidates include diamond,
which can have a refractive index as high as 2.7, \cite{edwards} and
transparent down to a wavelength of about 230 nm \cite{clark}, and
coherently-prepared atoms (confined in, say, a dielectric box) with a
resonantly-enhanced refractive index \cite{scully}, which can
theoretically reach the order of one hundred \cite{fleischhauer} and a
proof-of-concept experiment of which has already been demonstrated
\cite{zibrov}.

\begin{figure}[htbp]
\centerline{\includegraphics[width=0.48\textwidth]{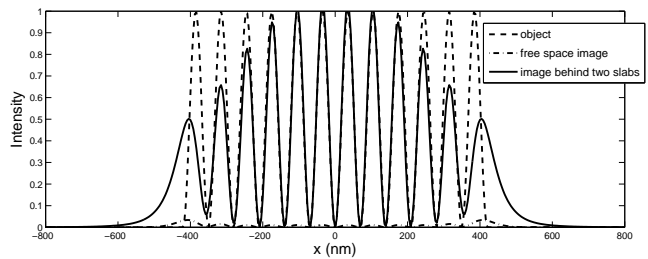}}
\caption{Plots of image intensities transmitted in free space and
by two dielectric slabs respectively.}
\label{lines}
\end{figure}

An outstanding problem of using any waveguide, except
negative-refractive-index slabs, for EWA is that ideal enhancement
only occurs for single-waveguide modes, which are discrete and
band-limited for each $\omega$. For instance, the discrete $k_x$'s of
the TE modes in a symmetric slab waveguide are determined by
Eq.~(\ref{pole}) and band-limited by Eq.~(\ref{resolutionlimit}).  As
a result, an object with frequency components that lie outside the
waveguiding band or do not coincide with the bound states cannot be
perfectly reproduced. Waveguide slabs are therefore particularly
suited to periodic image transmission.  As a simple example, consider
an approximately periodic pattern $\mathbf{E}_{0+}(z=0,x) =[0,1,
0]\sin(2\pi x/\Lambda)\textrm{rect}[x/(6\Lambda)]$ in a dielectric
medium with an index $n_1$ for $z\le 0$ to represent the source, and
at the object plane a dielectric-air interface is present at $z = 0$
to create evanescent waves for $z> 0$ via total internal reflection.
Assume that $\Lambda = 140$ nm, $\lambda = 230$ nm, $d = 2u= 2v = 20$
nm, $n_0 = 1$, $n_1 = 2.7$, and $a = 20$ nm, for a total transmission
distance of $80$ nm, and another dielectric medium with an index $n_1$
is present behind the image plane to convert the evanescent waves back
to propagating waves to represent the detector. The image intensities
in free space and behind two dielectric slabs, respectively, are
plotted in Fig.~\ref{lines}. The image enhancement due to the presence
of the dielectric slabs is clearly evident. One may also use a
multimode waveguide or a broadband light source to increase the amount
of transmitted spatial frequencies for more complex objects.

This work was sponsored by the Defense Advanced
Research Projects Agency (DARPA) Center for Optofluidic Integration.

\section*{Erratum}
In a previous Letter \cite{rewa}, we assert in Eq.~(8) that the
reflection coefficient of a double-waveguide structure, $R$, goes to
zero in the limit of single-waveguide resonance, Eq.~(5) in
Ref.~\onlinecite{rewa}. A more careful derivation shows that this is
not the case,
\begin{align}
&\quad \lim_{r'^2\exp(2ik_z'a)\to 1}R 
\nonumber\\&= 
\lim_{r'^2\exp(2ik_z'a)\to 1}
\Gamma + \frac{\tau^2\Gamma\exp(2ik_zd)}{1-\Gamma^2\exp(2ik_zd)}
\\
&=\lim_{r'^2\exp(2ik_z'a)\to 1}
\Gamma\left(1-\frac{\tau^2}{\Gamma^2}\right)
\label{eq2}\\
&=2r-\frac{tt'}{r'}.
\label{eq3}
\end{align}
Although the expression $(1-\tau^2/\Gamma^2)$ in Eq.~(\ref{eq2})
approaches zero, $\Gamma$ approaches infinity, and an application of
L'Hopital's rule shows that $R$ is given by Eq.~(\ref{eq3}), not zero.
That said, the possibility of a vanishing $R$, thereby achieving
reflectionless evanescent-wave amplification, still exists. This
condition is obtained by assuming that $R$ goes to zero but
neglecting the case in which $\Gamma = 0$,
\begin{align}
R &=\Gamma + \frac{\tau^2\Gamma\exp(2ik_zd)}{1-\Gamma^2\exp(2ik_zd)} = 0,
\\
\tau^2 &= \Gamma^2-\exp(-2ik_zd),\label{eigenvalue}
\end{align}
the total transmission coefficient, $T$, becomes
\begin{align}
T &= \frac{\tau^2\exp(ik_zd)}{1-\Gamma^2\exp(2ik_zd)}\\
&=-\exp(-ik_zd),
\label{T}
\end{align}
which, coincidentally, is the same as Eq.~(7) in
Ref.~\onlinecite{rewa}.  Hence, reflectionless evanescent-wave
amplification can still be achieved, provided that
Eq.~(\ref{eigenvalue}) in this Erratum, which now depends on $d$, the
distance between the two waveguides, is satisfied.

\begin{figure}[htbp]
\centerline{\includegraphics[width=0.45\textwidth]{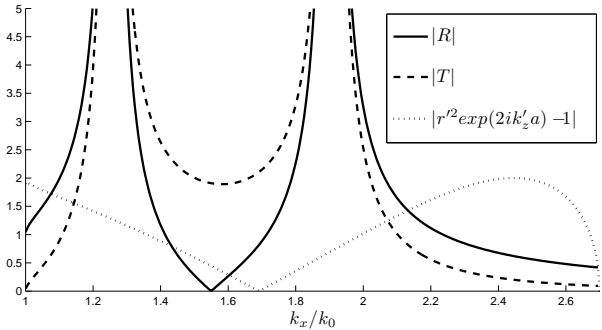}}
\caption{Plot of $|R|$, $|T|$ and $|r'^2\exp(2ik_z'a)-1|$ versus
$k_x/k_0$.}
\label{amplitude}
\end{figure}

Using the same example as in Ref.~\onlinecite{rewa}, where $\lambda =
230$ nm, $d = 20$ nm, $n_0 = 1$, $n_1 = 2.7$, and $a = 20$ nm,
Fig.~\ref{amplitude} plots $|R|$, $|T|$, and $|r'^2\exp(2ik_z'a)-1|$
versus $k_x/k_0$.  It is clear that $R$ is finite when
$r'^2\exp(2ik_z'a)-1$ is zero, but there exists a different $k_x$ at
which $R$ vanishes, while $|T|$ at this $k_x$ has the desired value
$\exp(-ik_z'd)\approx 1.92$ predicted by Eq.~(\ref{T}).
The double-peak shape of $R$ and $T$ is due to the non-degenerate
waveguide modes of the double-waveguide structure, which would create
the undesirable multiple evanescent-wave reflections as described in
Ref.~\onlinecite{rewa}.

The numerical example depicted by Fig.~2 in Ref.~\onlinecite{rewa},
calculated using a multiple-scattering analysis, is still correct and
unaffected by the above issue.

In conclusion, when single-waveguide resonance is reached, the
double-waveguide structure is capable of evanescent-wave
amplification, but not without reflection. Reflectionless
evanescent-wave amplification can still be achieved by two dielectric
planar waveguides, provided that Eq.~(\ref{eigenvalue}) in this
Erratum, not Eq.~(5) in Ref.~\onlinecite{rewa}, is satisfied.

\end{document}